\documentclass[]{pinchcr}
\usepackage{amsfonts}
\usepackage{amsmath}
\usepackage{amssymb}
\usepackage{graphicx}
\usepackage{epstopdf}

\begin{document}


\chapter[{\bf Dynamics of nano-magnetic oscillators}]{\bf Dynamics of nano-magnetic oscillators\\
{\normalsize T. Dunn,  A. L. Chudnovskiy, A. Kamenev }}

\section{Introduction}
\label{sec:intro}

The use of spin polarized currents for  manipulation of
nano-magnetic structures, known as {\em spintronics}, has grown
since it's first inception by Slonczewski and Berger
\shortcite{Berger96,Slonczewski96} into a thriving field of
research
\shortcite{Waintal00,Tserkovnyak02,Slonczewski-Sun,Ralph08,Tserkovnyak09,Chudnovskiy08,Bertotti,Ralph-Stiles}.
It has shown tremendous practical potential in it's ability to
quickly read and write non-volatile information into high density
memory structures by reversing the orientation of magnetic bits as
well as it's ability to drive magnetic oscillators, which have a
variety of uses
\shortcite{Burrowes10,Mizukami01,Tatara04,Li04,Parkin08,Tatara08,Beach08,Hals09,Halperin-Tserkovnyak,Silva08,Krivorotov2010}
.

Slonczewski and Berger predicted that a current of spin polarized electrons passing through a ferromagnetic layer generates a torque on the magnet and thus can be used to change the orientation of the magnetization. This effect is known as a spin transfer torque.
The spin-torque (ST) phenomenon consists of the transfer of angular momentum by the exchange interaction of electrons with the macroscopic magnetization.   Due to the conservation of the total angular momentum in the interaction between spin polarized current and the magnetization of the ferromagnet, the net change in the angular momentum for the current flowing through the magnetic layer must be equal to the net angular momentum absorbed by the ferromagnetic layer. In turn, the angular momentum is proportional to the magnetization, the coefficient being given by the gyromagnetic ratio, hence the absorption of the angular momentum results in the change of the magnetization direction, that is  in a torque on the ferromagnet. The existence of ST was confirmed experimentally  \shortcite{Myers99,Tsoi98}.  This phenomenon provides the physical mechanism for the operation of spintronics devices.

\begin{figure}[h]
  \begin{centering}
  \includegraphics[width=10cm]{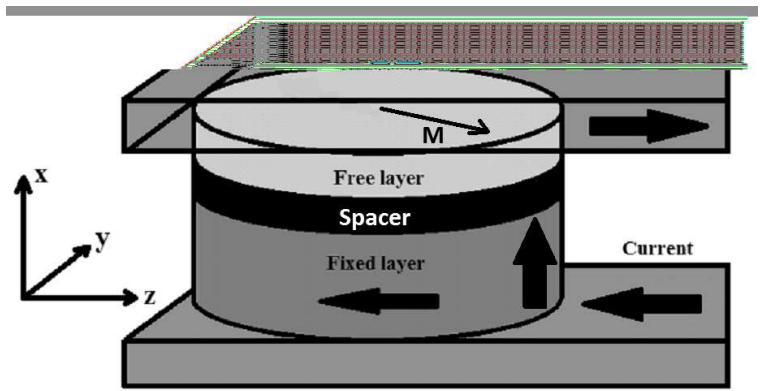}
  \par\end{centering}
  \caption{Schematic diagram of magnetic tunnel junction (MTJ). Electrons run clockwise through the device first passing through the bottom "fixed" layer, where they become partially polarized in the $-\mathbf{\hat{z}}$-direction. They then pass through the "free" layer imparting some angular momentum.  The MTJ has it's easy axis along the $\mathbf{\hat{z}}$-axis with the easy plane normal to the $\mathbf{\hat{x}}$-axis. }\label{fig:diagram}
\end{figure}

Magnetic nano-pilar  structures  typically consist of two magnetic layers, hereafter referred to  as the "free" and the "fixed" layer, and one non-magnetic layer (e.g. a tunneling-transparent insulator or a metal) that are sandwiched in a pillar between two ohmic contacts, Fig.~\ref{fig:diagram}. The fixed and free layers differ significantly in  their coercivity  fields, with the fixed layer being pinned along it's orientation much more strongly than the free layer. This can be done by increasing the thickness of the fixed layer, pinning it by proximity with an anti-ferromagnet, or picking a harder magnetic material. The free layer alternatively is much easier to manipulate. When an electric current is passed through the fixed layer in Fig.~\ref{fig:diagram} in the $\mathbf{\hat{x}}$-direction, it becomes polarized along the direction of the fixed layer. When it encounters the free layer the polarized spin-current induces ST and thus may change the magnetization direction of the layer, allowing for a number of dynamic regimes.

One such regime involves using ST to switch the orientation of the free layer in a magnetic tunnel junction (MTJ). This use of ST was confirmed shortly after it's theoretical prediction \shortcite{Katine00,Grollier01} with much of the early research focusing on understanding the effects of internal anisotropies \shortcite{Sun00,Mangin06}, temperature \shortcite{Ozyilmaz03,Urazhdin03,Hosomi05}, and spin-current strength \shortcite{Myers02,Seki06} on the switching dynamics. With the development of high tunneling magnetoresistance  MTJ's \shortcite{Yuasa07} and with switching times in the sub-nanosecond regime \shortcite{Tulapurkar04,Kent04} ST random access memory has become a very attractive candidate for non-volatile memory. While development of ST memory has moved forward \shortcite{Matsunaga09} recent research has also focused on more exotic switching strategies. Some of these strategies include using high frequency ac spin-currents to excite resonant spin modes to assist the switching \shortcite{Cui2008} as well as using spin-current polarized perpendicular to the interface \shortcite{Krivorotov_APL2010} to increase energy efficiency and reduce switching times \shortcite{Bedau10}.

Another major area of research for MTJs focusses on the use of MTJ's as small oscillators, known as spin-torque oscillators. Oscillations in MTJ's were first observed in some of the earliest studies of ST \shortcite{Kiselev03,Rippard04}. Since  ST oscillators can operate in GHz frequencies they've been an attractive candidate for use in timing mechanisms for computers and in radio-frequency devices for use in radar and telecommunication devices \shortcite{Deac08}.  In the last few years it has also been shown that ST oscillators can exhibit resonant excitations with ac spin currents which oscillate near their natural frequency \shortcite{Georges09,Ruotolo09,Krivorotov2010,Krivorotov_cm1004,Rippard10,Krivorotov_cm1103}.
Because of these practical applications, much of the research done in the last decade  has centered on the reduction and control of the oscillation linewidth \shortcite{Mizukami01,Krivorotov05,Nazarov06,Pribiag09}. While it's been shown that external fields \shortcite{Thadani08,Braganca10} and oscillation amplitude \shortcite{Rippard06,Kim07,Krivorotov-cm0812} can be used to tune the linewidth, the linewidth itself is the result of noise in the MTJ and much research has looked at understanding sources of noise and decay  in ST oscillators \shortcite{Ralph05,Mistral06,Slavin07}.
Thus understanding of stochastic processes such as thermal and shot noise has become increasingly important for operation of ST devices,
as manufacturing technology has improved and these devices have been pushed to smaller and smaller sizes.

First description of thermal noise in ferromagnets has been developed in the pioneering work by W. F. Brown \shortcite{Brown63}, where the noise have been described as a stochastic magnetic field acting on the magnetization. The  Fokker-Planck  equation for the probability distribution of magnetization direction has been formulated, and served as a basis for consideration of the noise in various dynamical regimes.  Further development of theoretical treatment of the noise has been provided by \shortciteNP{Tserkovnyak05,MacDonald,Foros08}. A crucial simplification of the theoretical description of the magnetization noise has been reached in the work of \shortciteNP{ApalkovPRB} by noticing that the dynamics of ST device exhibits a time scale separation between the fast magnetization precession and slow change in the energy of the precessional orbit. This allowed to average the equations of motion over the fast variables and formulate the one dimensional Fokker-Planck equation for the energy distribution.

In this chapter we explore how stochastic processes effect switching elements as well as ST oscillators. To do so we first discuss the deterministic dynamics of MTJs followed by the corresponding stochastic dynamics. Our particular emphasis is on the effects of out-of-equilibrium
noise, such as spin-torque shot-noise, inherent to any device operating under an applied current. Following
Ref.~\shortcite{ApalkovPRB}, we derive a stochastic Langevin equation for the slowly varying energy of precessional orbit, thereby introducing the energy noise and the energy diffusion coefficient. We then formulate a Fokker-Planck equation for the energy density distribution.
We use these constructions to analyze switching time distribution as well as the shape of the optimal spin-current pulse, which  minimizes Joule losses of a switch. We also derive generic expression for the linewidth of ST oscillator and discuss its dependence on temperature, spin-current amplitude and other parameters.

\section{Deterministic dynamics}
\label{sec:deterministic-motion}

In this section we briefly discuss magnetization dynamics  of a
mono-domain ferromagnet with the emphasis on the time scale
separation, which allows to introduce magnetic energy as a slow
(hydrodynamic) variable. On the most basic level the magnetization
${\bf M}$ precesses around the instantaneous effective magnetic
field ${\bf H}_{\mathrm{eff}}$ according to the Landau-Lifshitz
(LL) equation
\begin{equation}
{\bf\dot{M}}_{\mathrm{LL}}=-\gamma \big[ {\bf M}\times{\bf H}_{\mathrm{eff}}\big] ,
\label{eq:precession}
\end{equation}
where $\gamma$ is the gyromagnetic ratio. The effective magnetic
field is given by the gradient of the magnetic energy $E({\bf M})$
with respect to the magnetization, i.e. ${\bf H}_{\mathrm{eff}}=
-\nabla_{\bf M} E({\bf M})$. The  energy of magnetic
layers, conventionally used in magnetic junction devices,
typically includes  a combination of an easy plane and the
in-plane easy axis anisotropy as well as the effect of an
externally applied magnetic field
\begin{equation}
{E({\bf M})}= {\mu_0}\left[ -\frac{H^z_k}{2 M_\mathrm{s}} \left( \mathbf{M} \cdot
\mathbf{\hat{z}} \right)^2 +\frac{H^x_k}{2 M_\mathrm{s}} \left( \mathbf{M}
\cdot \mathbf{\hat{x}} \right)^2 -\mathbf{H}_{\mathrm{ext}}\cdot
\mathbf{M}\right] .
\label{eq:anisotropy}
\end{equation}
Here $H^z_k$ represents the strength of the easy axis anisotropy,
chosen in the $z$-direction, while $H_k^x$ represents the strength
of the easy $(y-z)$-plane anisotropy field,
Fig.~\ref{fig:diagram}. The anisotropy fields create an energy
landscape with two valleys separated by an energy barrier. The
minima of these valleys determine the orientation of the two
stable states. External fields $\mathbf{H}_{\mathrm{ext}}$ are often used to
force the magnetization to align in a specific direction or to
allow one of the energy minima to be lower than the other.

\begin{figure}[h]
  \begin{centering}
  \includegraphics[width=6cm]{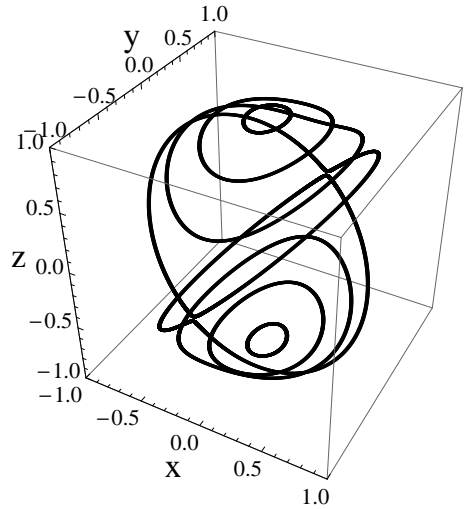}
  \par\end{centering}
  \caption{Several closed SW orbits of constant energy in the northern and southern hemispheres for $H_k^z = H_k^x = M_\mathrm{s}$. }\label{fig:SW-orbits}
\end{figure}

The Landau-Lifshitz equation (\ref{eq:precession}) evidently
conserves both the absolute value of the {\em total} magnetization $|{\bf
M}|=M_\mathrm{s}$ (subscript $\mathrm{s}$ stays for the {\em saturation} magnetization) and the magnetic energy $E({\bf M})$. As a
result, the magnetization is confined to move along a closed orbit
of a constant energy $E({\bf M})=E$, belonging to a sphere of radius $M_\mathrm{s}$. These are the so-called
Stoner-Wohlfarth (SW) orbits, plotted in
Fig.~\ref{fig:SW-orbits}.

It was realized long ago that the actual magnetization dynamics
must include the energy dissipation, originating from the coupling of
the global magnetization ${\bf M}$ with the host of microscopic
degrees of freedom. The simplest phenomenological way to
incorporate these effects in the magnetization dynamics was
suggested by Gilbert. The corresponding Gilbert damping (GD) term
in the  equation of motion  leads to the decay of the precession
and aligns the magnetization along the effective magnetic field
\begin{equation}
{\bf\dot{M}}_{\mathrm{GD}}= \frac{\alpha}{M_\mathrm{s}}\,  \big[{\bf M}\times{\bf\dot{M}}\big].
            \label{eq:Gilbert_damping}
\end{equation}
The strength of the Gilbert dissipation is proportional to the
phenomenological dimensionless damping constant $\alpha$. In
modern nano-magnetic devices its value may be as small as
$\alpha=0.01$ \shortcite{Myers99,Katine00,Ozyilmaz03}, allowing for dozens
precession cycles prior to equilibration.

Dramatic progress of the last decade is associated with the
realization by  Slonczewski \shortcite{Slonczewski96} and Berger
\shortcite{Berger96} that the magnetization may be influenced by
the spin polarized current of electrons through the ST effect.
Entering the free layer, spin polarized electrons found themselves either aligned (with the amplitude $\propto \cos\theta/2$, where $\theta$ is an angle between polarizations of the fixed and free layers), or antialigned (with the amplitude $\propto \sin\theta/2$) with the free layer magnetization direction. In the latter case the strong exchange interactions cause the rapid spin-flip, transferring the
angular momentum $\hbar$ from the itinerant electron to the macroscopic polarization of the free layer.
The corresponding non-conservative term in the macroscopic equation of motion takes the form
\begin{equation}
{\bf\dot{M}}_{\mathrm{ST}}=\frac{\gamma}{M_\mathrm{s}^2}\,  \left[{\bf{M}}\times [
\bf{ \hat{\mathcal{I}}_s}\times \bf{ M}]
\right],
\label{eq:spin-torque}
\end{equation}
which tends to align the magnetization along the spin polarization
of the electric current.
The spin-current vector $\,\bf{ \hat{\mathcal{I}}_s}\, $ is directed along spin polarization of the fixed layer, Fig. \ref{fig:diagram}. Its absolute value is proportional to the
difference of the electric currents of spin-up and spin-down
electrons, $\, \mathbf{\mathcal{I}_s}=\left(I_{\uparrow}-I_{\downarrow}\right)\hbar/2e$.

The three above mentioned terms provide a conventional deterministic description of the magnetization dynamics
\begin{equation}
{\bf\dot{M}}={\bf\dot{M}}_{\mathrm{LL}}+{\bf\dot{M}}_{\mathrm{GD}}
+ {\bf\dot{M}}_{\mathrm{ST}}.
\label{eq:LLGS}
\end{equation}
\begin{figure}[h]
  \begin{centering}
  \includegraphics[width=4.5cm]{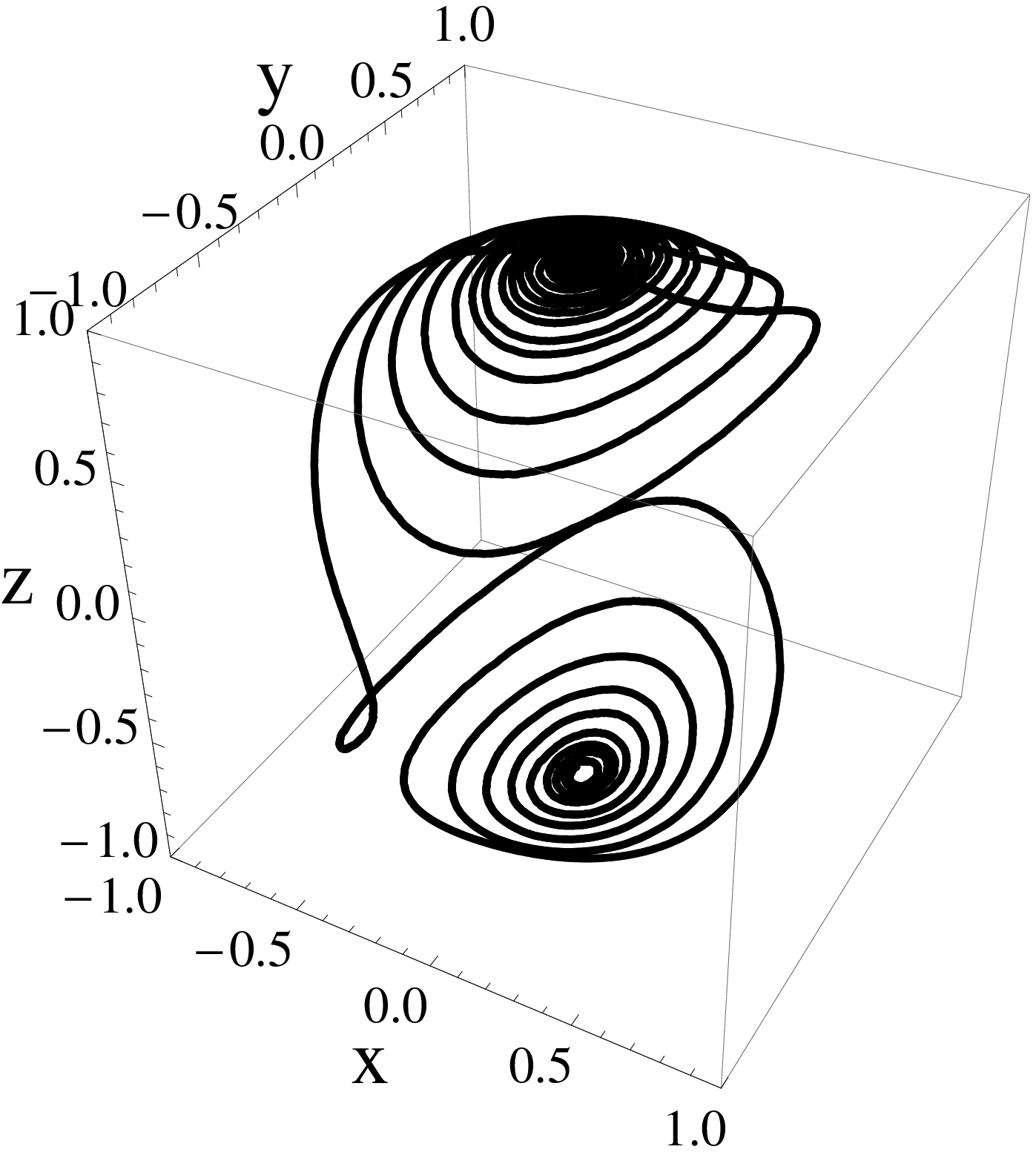}
  \includegraphics[width=8cm]{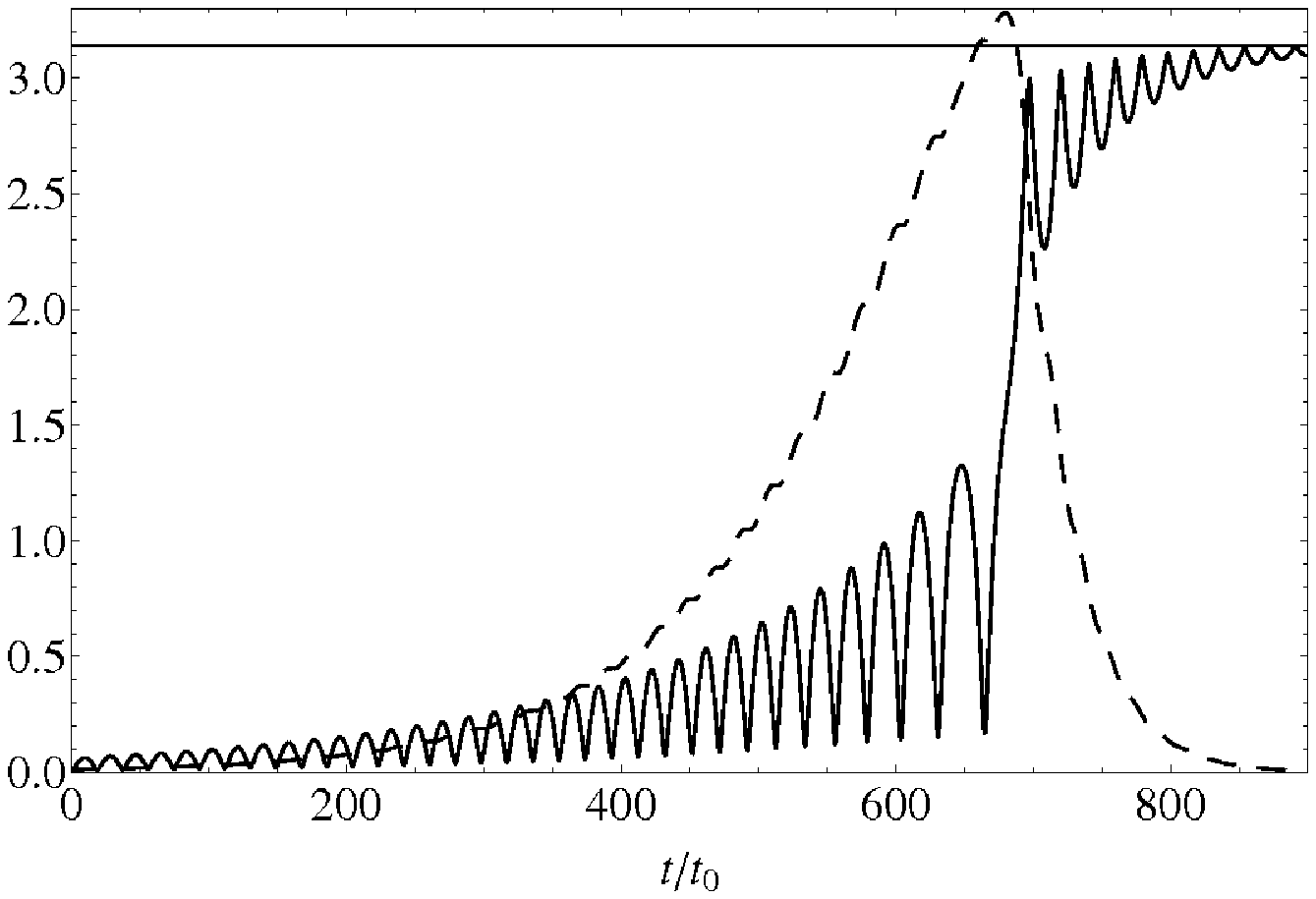}
  \par\end{centering}
  \caption{(Left) Switching trajectory for $H^z_k = H^x_k = M_\mathrm{s}$ and
  $\mathcal{I}_s = -0.05 M_\mathrm{s}$.  (Right) Azimuthal angle $\theta$ vs. time (full line) and energy vs. time (dashed line)
  for magnetization reversal of the system with $\alpha = 0.01$, $H^z_k = 0.028 M_\mathrm{s}$, $H^x_k = M_\mathrm{s}$,
  and $\mathcal{I}_s = -0.01 M_\mathrm{s}$. Time is measured in units of $t_0 = (\gamma M_\mathrm{s})^{-1}$
  with energy in units of $\pi E_0 = \mu_0 H_k^z M_\mathrm{s} /2$.}
  \label{fig:switchprocess}
\end{figure}
Fig.~\ref{fig:switchprocess} shows magnetization vector as well as the azimuthal
angle $\theta$ and magnetic energy (\ref{eq:anisotropy})  as functions of time   in the regime corresponding to the current-induced
magnetization switching experiment.  In this case the spin current is
sufficiently strong to overcome the
damping and make magnetization to cross the energy barrier, eventually coming to rest
along negative $\bf \hat{z}$-axis. As
Fig.~\ref{fig:switchprocess} shows the magnetization precesses
around the $\bf \hat{z}$-axis many times before crossing over the
energy barrier. It may be noticed that, while the angle oscillates widely on the time scale of the precession period,
the energy is a smooth function, which changes only at a much longer time scale.
The reason for this behavior is that the energy-conserving LL term in eqn~(\ref{eq:LLGS}) is typically much larger
than energy-non-conserving dissipative GD as well as driving ST terms.

This observation suggests to parameterize the instantaneous
magnetization direction by the slow energy variable $E$ and a fast
$2\pi$-periodic variable $\varphi$, which runs uniformly along a
closed SW orbit of a constant energy.  Up to precession period
$P_E$ the latter is simply the time needed to reach a given point
along SW orbit $d\varphi =2\pi dt/P_E$, where according to
Eq.~(\ref{eq:precession})
\begin{equation}
dt=\frac{d{\bf M}\cdot \left[{\bf H}_{\mathrm{eff}}\times{\bf M}\right]}{\gamma \left|[{\bf H}_{\mathrm{eff}} \times{\bf M}]\right|^2}\,;
\quad\quad\quad
P_E=\oint \frac{d{\bf M}\cdot \left[{\bf H}_{\mathrm{eff}}\times{\bf M}\right]}{\gamma\left|[{\bf H}_{\mathrm{eff}} \times{\bf M}]\right|^2}\, .
\label{P_E}
\end{equation}
Here the integral runs along the orbit of energy $E$.
We thus notice that the parametrization ${\bf M} ={\bf M}(E,\varphi)$ is specified implicitly by the following relations:
 \begin{equation}
\partial_\varphi{\bf M} = \frac{\gamma}{\Omega_E}\, [{\bf H}_\mathrm{eff}\times {\bf M}]\,; \quad\quad\quad\quad
\partial_E   {\bf M} = -\frac{\big[ {\bf M }\times [{\bf H}_\mathrm{eff}\times {\bf M}]\big]}
{\big|[{\bf H}_\mathrm{eff}\times {\bf M}]\big|^2} \,,
                                                                  \label{eq:implicit-derivatives}
\end{equation}
where $\Omega_E=2\pi/P_E$ is the precession frequency. The first
one is a consequence of LL equation (\ref{eq:precession}) and our
choice of $\varphi$, while the second can be checked by e.g. going
to the instantaneous reference frame, where ${\bf H}_\mathrm{eff}$
is directed along the $z$-axis. Notice that $\partial_\varphi{\bf
M}\cdot \partial_E   {\bf M}=0$ and thus the new coordinates are
locally orthogonal. Such parameterization can't be global, of
course, but rather should be introduced separately for each of the
{\em four} basins  apparent in Fig.~\ref{fig:SW-orbits}. Notice
also that
 \begin{equation}
\frac{d \mathcal{A}_E}{d E} =  \oint \frac{{\bf M} \cdot
[\partial_E {\bf M}\times d{\bf M}]}{M_\mathrm{s}}  = \oint
\frac{[{\bf M} \times  \partial_E {\bf M}]\cdot d{\bf
M}}{M_\mathrm{s}} = \gamma M_\mathrm{s} P_E \,,
                                                                  \label{eq:area-period}
\end{equation}
where $\mathcal{A}_E$ is the oriented area of the sphere, enclosed
by the orbit with energy $E$. Here we employed
eqs~(\ref{eq:implicit-derivatives}) and (\ref{P_E}). Equation
(\ref{eq:area-period}) provides geometrical interpretation of the
period as a change of the orbit's area.

One can write now the equation of motion (\ref{eq:LLGS}) as equations for $E(t)$ and $\varphi(t)$. To this end we notice that
\begin{equation}
\dot{E}=\nabla_{\bf M} E \cdot {\bf \dot{M}}=-{\bf H}_\mathrm{eff}\cdot {\bf \dot{M}}\,; \quad\quad\quad\quad
 \dot{\varphi}=\frac{\Omega_E}{\gamma}\,
 \frac{\left[{\bf H}_{\mathrm{eff}}\times{\bf M}\right]\cdot {\bf \dot M}}
 {\left|[{\bf H}_{\mathrm{eff}} \times{\bf M}]\right|^2}.
                                                                                                               \label{dot_E}
\end{equation}
and substitute ${\bf \dot{M}}$ from eqn~(\ref{eq:LLGS}). The fast LL part obviously drops from the equation for the energy, while the remaining
slow terms acquire the form
\begin{equation}
\dot{E} = -  F(E,\varphi) + {\bf \hat{\mathcal{I}}_s}  \cdot {\bf V} (E,\varphi)\, .
\label{eq:E_det0}
\end{equation}
The two {\em generalized forces} on the right hand side originate from the Gilbert damping  and spin-torque
\begin{equation}
F(E,\varphi) =\frac{\alpha}{M_\mathrm{s}} \, \left[\dot{\mathbf{M}}\times \mathbf{H}_{\mathrm{eff}}\right] \cdot \mathbf{M}\,;
\quad\quad \quad
{\bf V}(E,\varphi) = \frac{1}{M_\mathrm{s}^2 } \left[ \dot{\mathbf{M}}\times \mathbf{M}\right].
\label{eq:I_E}
\end{equation}
correspondingly. Hereafter $ \dot{\mathbf{M}}$ is understood as a conservative LL part, eqn~(\ref{eq:precession}), only. On the other hand,
Gilbert damping drops from the equation for the angular variable $\varphi$ and its dynamics is mostly governed by the uniform precession
\begin{equation}
\dot{\varphi}=\Omega_E+\Omega_E\frac{\mathcal{I}_{\mathrm{s}}\cdot\left[{\bf M}\times{\bf H}_{\mathrm{eff}}\right]}{\left|[{\bf H}_{\mathrm{eff}} \times{\bf M}]\right|^2} .
\label{eq_varphi_det}
\end{equation}
The second term here originates from the fact that the spin-torque may have a component along the SW orbit. This leads to
a certain renormalization of the precession frequency $\Omega_E$. Since the integral along the closed orbit of the second term on the right hand side of eqn~(\ref{eq_varphi_det}) vanishes, such renormalization is of the order $O(\mathcal{I}_{\mathrm{s}}^2)$. In the spirit of time scale separation we shall neglect this kind of corrections. Similarly we disregard all terms which are of the order $O(\alpha^2)$, assuming a small Gilbert damping constant.

The above remarks show that the fast variable $\varphi$
approximately exhibits the uniform precession according to
$\dot{\varphi}=\Omega_E=2\pi\gamma M_\mathrm{s} dE/d\mathcal{A}$,
cf.~eqn~(\ref{eq:area-period}), allowing to identify
$\mathcal{A}/2\pi\gamma M_\mathrm{s}$ with the action canonically
conjugated to the angle $\varphi$. It suggests to average equation
(\ref{eq:E_det0}) for the slow energy variable over the precession
period, arriving at
\begin{equation}
 \dot{{E}}=-  {F}_E + {\bf \hat{\mathcal{I}}_s}  \cdot {{\bf V}}_E
\label{eq:energy-rate}
\end{equation}
where the averages of the generalized forces are given by
\begin{equation}
{F}_E =\frac{1}{M_\mathrm{s}P_E} \oint \alpha\, \left[d\mathbf{M}\times \mathbf{H}_{\mathrm{eff}}\right] \cdot \mathbf{M}\,;
\quad\quad \quad
{{\bf V}}_E = \frac{1}{M_\mathrm{s}^2 P_E}\oint \left[ d\mathbf{M}\times \mathbf{M}\right]
\label{eq:I_E-averaged}
\end{equation}
correspondingly. The integrals here run along SW orbit with energy
$E$.  Notice that to evaluate these forces one does not need any
information about dynamics, but rather only the form of {\em
static} SW orbits. While this strategy was shown useful in the
analysis of non-linear oscillators long ago \shortcite{Dykman79}
it was probably first applied in the present context  by Apalkov
and Visscher \shortcite{ApalkovPRB}  and further developed by
Bertotti \shortcite{Bertotti}.

\begin{figure}[h]
  \begin{centering}
  \includegraphics[width=10cm]{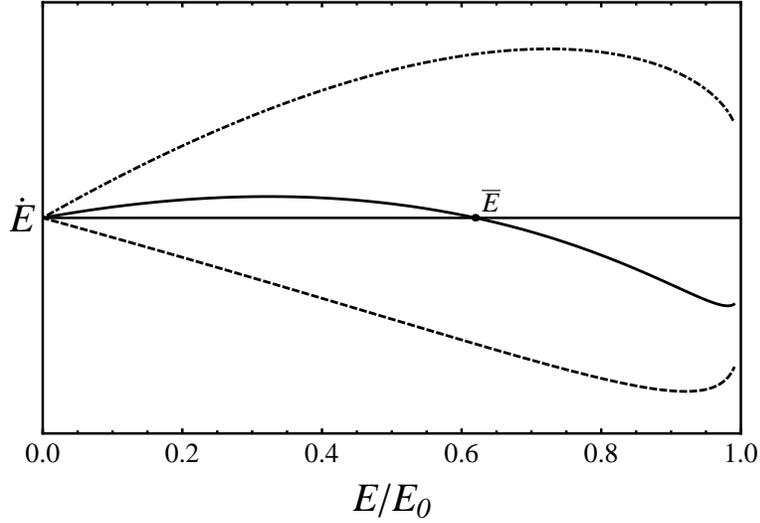}
  \par\end{centering}
  \caption{Shows the generalized force, $\dot{{E}} = -  {F}_E + {\bf \hat{\mathcal{I}}_s}  \cdot {\bf V}_E$ vs.
   energy for various strengths of $\mathcal{I}_s$ for system with $H_k^z = 0.028 M_\mathrm{s}$ and $H_k^x = M_\mathrm{s}$.
   From top to bottom $\mathcal{I}_s=0.008 M_\mathrm{s}$ (dot-dashed), $\mathcal{I}_s = 0.006 M_\mathrm{s}$ (solid), $\mathcal{I}_s = 0.004 M_\mathrm{s}$ (dashed).}
   \label{fig:energycurrent}
\end{figure}

Figure \ref{fig:energycurrent} shows the total generalized force as a function of energy for
various values of the spin-current $\,\mathcal{I}_s$.  For sufficiently small spin-current
the force is strictly negative (dashed line, $\dot{E}<0$) forcing the magnetization to relax to the
equilibrium energy minimum.   At a larger current there is an energy value $\bar E$, where $\dot { E}=0$, corresponding to
a stable  precession around SW orbit with the energy $\bar E$ \shortcite{Kiselev03,Mistral06}. Finally at a spin current larger than a certain
critical value the generalized force is positive in the entire energy interval (dot-dashed  line, $\dot{E}>0$). As a result the magnetization
is forced to switch to the new stable position. In the following sections we examine effects of noise on both switching and stable precession.

\section{Equilibrium stochastic dynamics of magnetization}
\label{sec:stochastic}

The deterministic description of the previous section is incomplete, especially for small enough magnetic domains. The reason
is the stochastic part of the magnetic torque. The fact that noise must accompany  Gilbert damping to satisfy the equilibrium
fluctuation-dissipation theorem (FDT) was first realized by W. F. Brown~\shortcite{Brown63}. The way he introduced the noise was by adding a
stochastic component to the effective magnetic field in eqn~(\ref{eq:precession}), ${\bf H}_{\mathrm{eff}}\to {\bf H}_{\mathrm{eff}}+{\bf h}(t)$,
where ${\bf h}(t)$ is an isotropic Gaussian noise. Its amplitude may be uniquely determined from FDT and is given by
\begin{equation}
\langle  {\bf h}^i(t) {\bf h}^j(t') \rangle =2\delta_{ij}\delta(t-t')\alpha k_\mathrm{B} T/M_\mathrm{s} \gamma=2\delta_{ij}\delta(t-t')D,
\label{h_noise}
\end{equation}
where $T$ is the temperature and $D$ is diffusion constant. In presence of the noise, the magnetization dynamics should be described by a probability distribution of the magnetization vector $\mathcal{P}({\bf M}, t)$, obeying a proper Fokker-Planck (FP) equation.

We shall first outline how it works in an equilibrium setting, generalizing it to the presence of the non-equilibrium spin-current in the next section.
Stochastic magnetic field adds a Langevin force term $\dot{\bf M}_{\mathrm{stoch}}=\gamma \left[{\bf h}(t)\times {\bf M}\right]$ to the equation of motion (\ref{eq:LLGS}) for the magnetization.   It leads to  Langevin force terms in equations of motion (\ref{eq:E_det0}), (\ref{eq_varphi_det}) for the energy-angle variables. The explicit form of the Langevin forces can be obtained using the projection  on $\varphi$- and $E$-directions respectively, as it was done for deterministic eqs  (\ref{dot_E}). As a  result we obtain the stochastic equations of motions with the {\em multiplicative} noise
\begin{eqnarray}
&&
\dot{E} = - F(E,\varphi) + {\bf g}_E(E,\varphi) \cdot{\bf h}(t)\,;
 \label{E_stoch}  \\
&&
\dot{\varphi}=\Omega_E+ {\bf g}_{\varphi}(E,\varphi) \cdot{\bf h}(t)\,,
\label{phi_stoch}
\end{eqnarray}
where the two mutually orthogonal noise-multiplying vectors are given by
\begin{equation}
{\bf g}_E=\dot{\bf M}=\Omega_E \partial_\varphi {\bf M}\,;
                                                                                                                            \label{g_E}
 \quad\quad\quad
{\bf g}_{\varphi}=\gamma\Omega_E [{\bf M}\times \dot{\bf M}]/|\dot{\bf M}|^2 = -\Omega_E\partial_E {\bf M}\,
\end{equation}
and we employed eqs~(\ref{eq:implicit-derivatives}). Since the multiplicative noise originates from the change of variables ${\bf M}\to E,\varphi$,
it should be treated with the Stratonovich regularization \shortcite{vankampen}. Employing the orthogonality $ {\bf g}_E\cdot {\bf g}_{\varphi}=0$,
one obtains the following FP equation for the probability density $ \mathcal{P}(E, \varphi, t)$:
\begin{equation}
\dot{\mathcal{P}}=\partial_E\left[ (F-k_E) \mathcal{P} +
D\partial_{E}\left( \left| {\bf g}_E \right|^2 \mathcal{P}\right) \right]
+ \partial_{\varphi}\left[(-\Omega_E-k_{\varphi}) \mathcal{P}
+ D\partial_{\varphi}\left(  \left|{\bf g}_\varphi \right|^2 \mathcal{P}\right)\right],
                                                                                                                        \label{Fokker-Planck}
\end{equation}
where the diffusion coefficient $D=\alpha k_\mathrm{B} T/M_\mathrm{s} \gamma$ is determined by the equilibrium correlator of stochastic magnetic field eqn (\ref{h_noise}). The ``false'' forces $k_E$ and $k_{\varphi}$ are given by
$k_{\mu}=D\sum_{\nu}(\partial_{\nu}{\bf g}_{\mu})\cdot {\bf g}_{\nu}$  \shortcite{vankampen,Kamenev11}, where $\mu,\nu=E,\varphi$. Employing
eqs (\ref{g_E}) along with orthogonality $\partial_\varphi{\bf M}\cdot \partial_E   {\bf M}=0$, one finds
\begin{eqnarray}
&&
k_E=   \frac{D}{2} \partial_E\left(\Omega_E^2 \left|\partial_{\varphi}{\bf M}\right|^2\right) +
\frac{D}{2} \Omega_E^2 \partial_E  \left|\partial_{\varphi}{\bf M}\right|^2 =
D(\partial_E-\partial_E \ln\Omega_E)  \left|  {\bf g}_E \right|^2 ;
                                                                                                \label{eq:k-E} \\
&&
k_{\varphi}=\frac{D}{2} \partial_\varphi\left(\Omega_E^2 \left|\partial_{E}{\bf M}\right|^2\right) +
\frac{D}{2} \Omega_E^2 \partial_\varphi  \left|\partial_{E}{\bf M}\right|^2 =
D\partial_\varphi  \left|  {\bf g}_\varphi \right|^2.
                                                                                                    \label{eq:k-phi}
\end{eqnarray}
This transforms the FP equation (\ref{Fokker-Planck}) into
\begin{equation}
\dot{\mathcal{P}}=\partial_E\left[ (F + {\mathcal D}_E \partial_E \ln\Omega_E) \mathcal{P} + {\mathcal D}_E \partial_{E} \mathcal{P} \right]
+ \partial_{\varphi}\left[-\Omega_E \mathcal{P} + {\mathcal D}_\varphi \partial_{\varphi} \mathcal{P}\right]\,,
                                                                                                                        \label{Fokker-Planck-1}
\end{equation}
where the two diffusion coefficients are ${\mathcal D}_\mu(E,\varphi) =D \left|  {\bf g}_\mu \right|^2$.

To proceed we employ the periodicity of  $\mathcal{P}(E, \varphi, t)$ in the angular directions to write it as a Fourier series
$\mathcal{P}(E, \varphi, t)=\sum_{m=-\infty}^{\infty}\mathcal{P}_m(E, t) e^{i m \varphi}$. One expects that on the fast time scale of the precession
period $P_E$ the distribution nearly equilibrates in the angular direction and then evolves slowly along the energy direction. This implies
that at long times $t\gg P_E$ one has $\mathcal{P}_0\gg \mathcal{P}_m$, where $m\neq 0$. As a result one can write a closed FP equation for
$\mathcal{P}_0(E,t)$ in the following form
\begin{equation}
\dot{\mathcal{P}}_0=\partial_E\left[ ({F}_E +\mathcal{D}_E \partial_E \ln\Omega_E) \mathcal{P}_{0} +
\mathcal{D}_E \partial_{E} \mathcal{P}_{0}\right]\,,
\label{Fokker-Planck_0}
\end{equation}
where the average generalized force $F_E$ is given by eqn~(\ref{eq:I_E-averaged}) and the averaged over the period energy diffusion
coefficient is
\begin{equation}
\mathcal{D}_E= \frac{1}{2\pi}\int_0^{2\pi}\!\!\! D\Omega_E^2 \left|\partial_{\varphi}{\bf M}\right|^2 d\varphi
= \frac{1}{P_E}\oint\!\! D \, \dot{\bf M} \cdot d{\bf M} = \frac{\gamma}{P_E}\oint\!\! D\,  [d{\bf M}\times {\bf H}_\mathrm{eff}] \cdot {\bf M}.
                                                                                                                \label{D_E}
\end{equation}
One may now include $\mathcal{P}_{\pm 1}$, $ \mathcal{P}_{\pm 2}$, {\em etc.} as progressively small corrections and take into account how
the incomplete angular equilibration affects evolution in the energy direction. We shall not take this root here.

Notice that in equilibrium FDT dictates $D=\alpha k_\mathrm{B} T/M_\mathrm{s} \gamma$ and thus $\mathcal{D}_E=k_\mathrm{B} T F_E$, cf. egn~(\ref{eq:I_E-averaged}). As a result of this remarkable relation the equilibrium stationary solution of the 1D FP equation (\ref{Fokker-Planck_0}) takes the  universal form
\begin{equation}
 \mathcal{P}_0(E)=  \frac{1}{Z} \,\exp\left[-\frac{E-k_\mathrm{B} T\ln \Omega_E^{-1} }{k_\mathrm{B} T}\right] =
 \frac{P_E}{2\pi Z}\, \exp\left[-\frac{E }{k_\mathrm{B} T}\right]\,.
\label{Boltsmann}
\end{equation}
where $Z$ is a normalization factor. The very last term here is,
of course, the Boltzmann exponent. It is crucial that  the entropy
$S(E)=k_\mathrm{B}T \ln  \Omega_E^{-1}$ has emerged, which may be
traced back to the density of states along SW orbit given by
$P_E\propto d\mathcal{A}/dE$, cf. eqn (\ref{eq:area-period}).
Notice that it is the ``false'' force $k_E$, eqn~(\ref{eq:k-E}),
which is responsible for this fact.

\section{Shot-noise and non-equilibrium stochastic dynamics}
\label{sec:shot-noise}

The advent of spin-polarized currents put the operation of spintronics devices in far from equilibrium conditions. This not only influences
the deterministic motion through the ST term (\ref{eq:spin-torque}), but also changes the stochastic force. Unlike the equilibrium noise, where the noise is uniquely determined by FDT, the out-of-equilibrium noise is non-universal and depends on details of a specific setup.
Here we consider nonequilibrium noise in a spin valve device consisting of two magnetic layers separated by a tunneling barrier, the so-called magnetic tunnel junction (MTJ). The MTJ is modeled by the two itinerant ferromagnets, whose majority ($\sigma=+$) and minority ($\sigma=-$) bands
are described by the operators $c^{\dagger}_{k\sigma}, c_{k\sigma}$ for the fixed ferromagnet and
$ d^\dagger_{l\sigma}, d_{l\sigma}$ for the free layer.
The corresponding Hamiltonian consists of three parts describing isolated fixed and free layers along with the tunneling of electrons between them
\begin{equation}
H=H_{\mathrm{fixed}}+H_{\mathrm{free}}+H_{\mathrm{tun}}.
\label{Ham2}
\end{equation}
The fixed layer is described by a Fermi-liquid Hamiltonian
\begin{equation}
H_{\mathrm{fixed}}=\sum_{k,\sigma} \epsilon_{k\sigma} c^{\dagger}_{k\sigma} c_{k\sigma}.
\label{Hfixed}
\end{equation}
The local exchange interaction of itinerant electrons with magnetic momentd in the free layer can be written as   $J_0 \int d{\bf r} {\bf S}({\bf r})\cdot {\bf s}({\bf r})$, where $ {\bf S}({\bf r})={\bf M}({\bf r})/\gamma$ is the spin density of the magnetic layer and ${\bf s}({\bf r})$ denotes the spin density of itinerant electrons. We assume that the spin current flowing through the free layer does not disturbe its monodomain magnetic structure. In that case, the layer can be considered as a single large spin with the magnetic moment ${\bf M_{\mathrm s}}$.
The Hamiltonian of the free layer that accounts for the interactions of the itinerant electrons in the free layer with its {\em total} spin ${\bf S} = {\bf M_{\mathrm s}}/\gamma$ is given by
\begin{equation}
H_{\mathrm{free}}=    \sum_{l\sigma}\epsilon_{l}  d^\dagger_{l\sigma} d_{l\sigma} - J{\bf S}\cdot {\bf s}-\gamma  {\bf S}\cdot {\bf H}_\mathrm{eff} \,,
\label{Hfree}
\end{equation}
where ${\bf s}=\frac{1}{2}\sum_{l\sigma\sigma'} d^\dagger_{l\sigma} \vec\sigma_{\sigma\sigma'} d_{l\sigma'}$ is the spin of itinerant electrons and $J$ is  Heisenberg exchange interaction constant.
Finally, the tunneling term in the Hamiltonian is given by
\begin{equation}
H_{\mathrm{tun}}= \sum_{kl,\sigma\sigma'}
W_{kl}^{\sigma\sigma'} c^\dagger_{k\sigma} d_{l\sigma'} + h.c. .
\end{equation}
Here the spin indices of the operators $c^\dagger_{k\sigma}$ and $d_{l\sigma'}$ denote the spin projections along the magnetization directions in the fixed and in free layers respectively. Because of a finite angle between the two magnetizations the tunneling matrix elements become spin-dependent. They are given by  $W_{kl}^{\sigma\sigma'} =  \langle \sigma|\sigma'\rangle W$, where the spin-transformation matrix is $\langle \sigma|\sigma\rangle=
e^{-i\sigma\phi/2} \cos\theta/2\,$ and $\langle \sigma|\sigma'\rangle=
e^{i\sigma\phi/2} \sin\theta/2$.
The angles $(\theta, \phi)$ are the polar and azimuthal angles that denote the direction of the magnetization in the free layer in the reference frame with $z$ axis pointing in the direction of magnetization of the fixed layer.

The main source of the non-equilibrium noise is the discrete
nature of the angular momentum transfer.  Indeed, each electron
tunneling at a random time may undergo a spin-flip with a
probability depending on the mutual orientation of the
quantization axis. Since each event transfers exactly the unit of
angular momentum $\hbar$, the direction of an ensuing
magnetization rotation is  random due to the uncertainty principle
(i.e. non-commutativity of the three components of the angular
momentum operator).  Therefore the electron transport is
accompanied by the stochastic force acting on the magnetization.
Its nature is similar to the charge shot noise.  While the latter
is due to the discreteness of charge, the former is due to the
quantization of the angular momentum. We call it thus the spin
shot noise \shortcite{Tserkovnyak05,Chudnovskiy08,Dunn10}.

To introduce this noise into the semiclassical equation of motion we write the model specified by eqn (\ref{Ham2}) as a path integral on the Keldysh contour \shortcite{Chudnovskiy08,Dunn10,Kamenev11} and integrate out all fermionic degrees of freedom, while keeping the dynamics of the macroscopic spin ${\bf S}$. The integration is facilitated by the tunneling approximation, i.e. keeping only the lowest nonvanishing terms in the coupling $W$. As a result the spin torque and spin shot noise terms as well as an additive renormalization of the Gilbert damping are generated. Furthermore,   the spin shot noise can be cast into the form of random fluctuations of the spin current vector $\delta{\bf I}_s$ entering the spin torque term (\ref{eq:spin-torque}). The noise correlator can be expressed in terms of MTJ conductance  for parallel $G_P$ and antiparallel $G_{AP}$ magnetization orientation of the two layers, ($G_P\geq G_{AP}$)
\begin{equation}\label{eq:correlator}
\left\langle \delta {\mathrm I}_\mathrm{s}^i(t) \delta {\mathrm I}_\mathrm{s}^j(t') \right\rangle =   2M_\mathrm{s}^2  D(\theta)\,\delta_{ij}\,\delta(t-t')\, ,
\end{equation}
where
\begin{equation}
D(\theta) =  \frac{ \alpha_0}{M_\mathrm{s}\gamma}\, k_\mathrm{B} T + \frac{\hbar}{2M_\mathrm{s}^2}\, {\mathrm I}_{\mathrm{sf}}(\theta) \coth\left(\frac{eV}{2k_{\mathrm{B}}T}\right)= D({\bf M}),
\label{eq:DI}
\end{equation}
where $\alpha_0$ is a bare Gilbert damping of an isolated layer  and $V$ is a voltage bias between the two ferromagnets.  The first term in eqn (\ref{eq:DI}) describes the equilibrium noise due to intrinsic relaxation processes in the active layer, it is therefore proportional to the intrinsic Gilbert damping constant $\alpha_0$.  The second term describes the additional noise due to the coupling of the active layer to the fixed layer which in this case plays the role of the external reservoir of spins. The external contribution to the noise is proportional to the spin-flip current ${\mathrm I}_{\mathrm{sf}}(\theta)$. The latter counts the {\em total} number of spin flips irrespective of the direction of the ensuing magnetization change  (as opposed to the spin current ${\mathcal I}_s$).  In the magnetic tunnel junction setup we found for the spin-flip conductance
\begin{equation}\label{eq:I_sf}
    \frac{d {\mathrm I}_{\mathrm{sf}}(\theta)}{d V}  = \frac{\hbar}{4e} \left[
    G_{P} \sin^2\!\left(\frac{\theta}{2}\right)+ G_{AP} \cos^2\!\left(
    \frac{\theta}{2}\right)\right] ;
\end{equation}
$$
G_P=G_{++}+G_{--}\,; \quad G_{AP} =G_{+-}+G_{-+}\, ,
$$
where we adopted notations of \shortcite{Slonczewski-Sun} for the partial conductances $G_{\sigma\sigma'}$  between
the spin-polarized bands of the two ferromagnets.
The external spin noise is accompanied by the renormalization of the Gilbert damping constant, which now depends on the
instantaneous magnetization of the free layer  through the angle $\theta$ it forms with the magnetization of the fixed layer
\begin{equation}
                                                                                              \label{eq:alpha-renorm}
     \alpha(\theta) = \alpha_0 +\frac{ \hbar \gamma}{ e M_\mathrm{s}}
   \left( \frac{d {\rm I}_{\mathrm{sf}}(\theta)}{d V} \right) = \alpha({\bf M}) \, ,
\end{equation}
where the spin-flip {\em differential} conductance is given by eqn~(\ref{eq:I_sf}).
For low voltages, $e V < k_{\mathrm{B}}T$, the tunneling contribution to the noise is proportional to the temperature and the total noise satisfies FDT with the renormalized Gilbert damping constant $\alpha({\bf M})$, \shortcite{Tserkovnyak05,Chudnovskiy08}. For voltages exceeding the temperature $e V > k_{\mathrm{B}}T$, the external noise is essentially nonequilibrium, proportional to the applied voltage and independent of temperature.

The conductances $G_{\sigma\sigma'}$, $G_P$, and $G_{AP}$ allow the complete characterization of the electric and magnetic properties of MTJ. So,
the electric conductance of the MTJ in an arbitrary orientation is given by
 $d{\mathrm I}_e(\theta)/dV=G_{P}\cos^2(\theta/2) + G_{AP}\sin^2(\theta/2)$. Notice that the spin shot noise is minimal for $P$ orientation and maximal for $AP$ one -- exactly opposite to the charge current and the charge shot noise.
In contrast to the spin-flip current introduced above, the spin-current $\mathcal{I}_s$ is governed by the spin conductance  \shortcite{Slonczewski-Sun}
\begin{equation}\label{I_s}
   \frac{ d{\mathcal I}_s}{dV} = \frac{\hbar}{4e}\left( G_{++}-G_{-+}+G_{+-}-G_{--}\right). \,
\end{equation}

Taking noise into account, one arrives at the stochastic equation of motion for magnetization
\begin{equation}
{\bf \dot M} = -\gamma \left[{\bf M}\times {\bf H}_{\mathrm{eff}}\right]+ \frac{\alpha(\theta)}{M_\mathrm{s}}  \left[{\bf M}  \times
{\bf \dot M}  \right]
+ \frac{\gamma}{M_\mathrm{s}^2}\,
\left[{\bf M}\times \left[\left({\bf \hat{\mathcal{I}}_s}+{\bf \delta  I}_\mathrm{s}(t)\right)   \times {\bf M}\right]\right] \,.
\label{eq:LLG}
\end{equation}
The equivalence of the two forms of description of the magnetization noise, either as a stochastic spin current or as a stochastic magnetic field, is expressed by the relation of the corresponding cumulants
\begin{equation}
\left\langle {\bf h}^i(t) {\bf h}^j(t')\right\rangle=\frac{1}{ M_\mathrm{s}^2 } \left\langle \delta {\bf I}_\mathrm{s}^i(t) \delta {\bf  I}_\mathrm{s}^j(t') \right\rangle = 2  D({\bf M})\,\delta_{ij}\,\delta(t-t')\,.
\label{eq:corr_h}
\end{equation}
Notice that the noise correlator  itself (\ref{eq:DI}) as well as
the renormalized Gilbert damping (\ref{eq:alpha-renorm})    depend
on the instantaneous angle between the free and fixed layers
magnetization direction. Being derived from the path-integral
representation, the correlator (\ref{eq:corr_h}) should be
regularized in the retarded (Ito) way.  Moreover, in presence of
the spin-torque the Langevin equation (\ref{E_stoch}) for the
energy variable acquires the additional deterministic term ${\bf
\hat{\mathcal{I}}_s} \cdot {\bf V} (E,\varphi)$, see
eqs~(\ref{eq:energy-rate}) and (\ref{eq:I_E-averaged}). This term
describes  pumping of the energy into the magnetic system by the
spin torque process. Together with the Gilbert damping,  the spin
torque  term determines the dynamics of the spin switching process
as well as the average energy of the SW orbit for the steady state
magnetization precession.

One can now generalize derivation of FP equation (\ref{Fokker-Planck_0}), presented in section \ref{sec:stochastic}, for the presence of the spin-torque and non-equilibrium noise. The result is given by the continuity relation in the energy direction
$\dot{\mathcal{P}}_0 +\partial_E \mathcal{J}_0=0$, where the probability current in the energy direction is given by
\begin{equation}
\mathcal{J}_0(E,t) = -\left[{F}_E - {\bf \hat{\mathcal{I}}_s}  \cdot {\bf V}_E + \mathcal{D}_E \partial_E \ln\Omega_E\right]  \mathcal{P}_{0} -
\mathcal{D}_E \partial_E \mathcal{P}_{0}\,.
\label{eq:FP_general-current}
\end{equation}
Here the components of the deterministic force and the energy diffusion coefficient are given by
eqs~(\ref{eq:I_E-averaged}) and (\ref{D_E}), correspondingly, where one has to substitute $\alpha\to \alpha({\bf M})$ and
$D\to D({\bf M})$ in accordance with eqs~(\ref{eq:alpha-renorm}) and (\ref{eq:DI}).

The stationary solution of FP equation (\ref{eq:FP_general-current})  plays a special role in several regimes, such as activated magnetization switching \shortcite{ApalkovPRB} or steady state magnetization precession \shortcite{Kiselev03},  considered below in more details. This solution is obtained by putting $\mathcal{J}_0 = 0$ in eqn (\ref{eq:FP_general-current})  and reads as
\begin{equation}
\mathcal{P}_0(E) \propto P_E \exp\left[-\int^E\frac{ F_E - {\bf \hat{\mathcal{I}}_s}  \cdot {\mathbf{V}}_E}{\mathcal{D}_E}\,  dE\right]\,.
\label{eq:stat}
\end{equation}
Notice that in equilibrium, i.e. ${\bf \hat{\mathcal{I}}_s}=0$ and $D({\bf M})=\alpha({\bf M})k_\mathrm{B}T/M_\mathrm{s}\gamma$, one has $\mathcal{D}_E = k_\mathrm{B}T {F}_E$, even for the magnetization-dependent damping constant and the noise correlator! As a result the Boltzmann form (\ref{Boltsmann}) with the entropic factor given by $P_E$ still holds. However, away from  equilibrium the stationary energy distribution (\ref{eq:stat}) may be rather different from the Boltzmann shape. The 1D Fokker-Planck equation  (\ref{eq:FP_general-current}) is especially useful to analyze non-stationary regimes of spin-valves, such as e.g. kinetics of the switching process.

\section{Dynamics and optimization of the magnetization switching}
\label{sec:switching-dynamics}

\begin{figure}
\begin{centering}
\includegraphics[width=6cm]{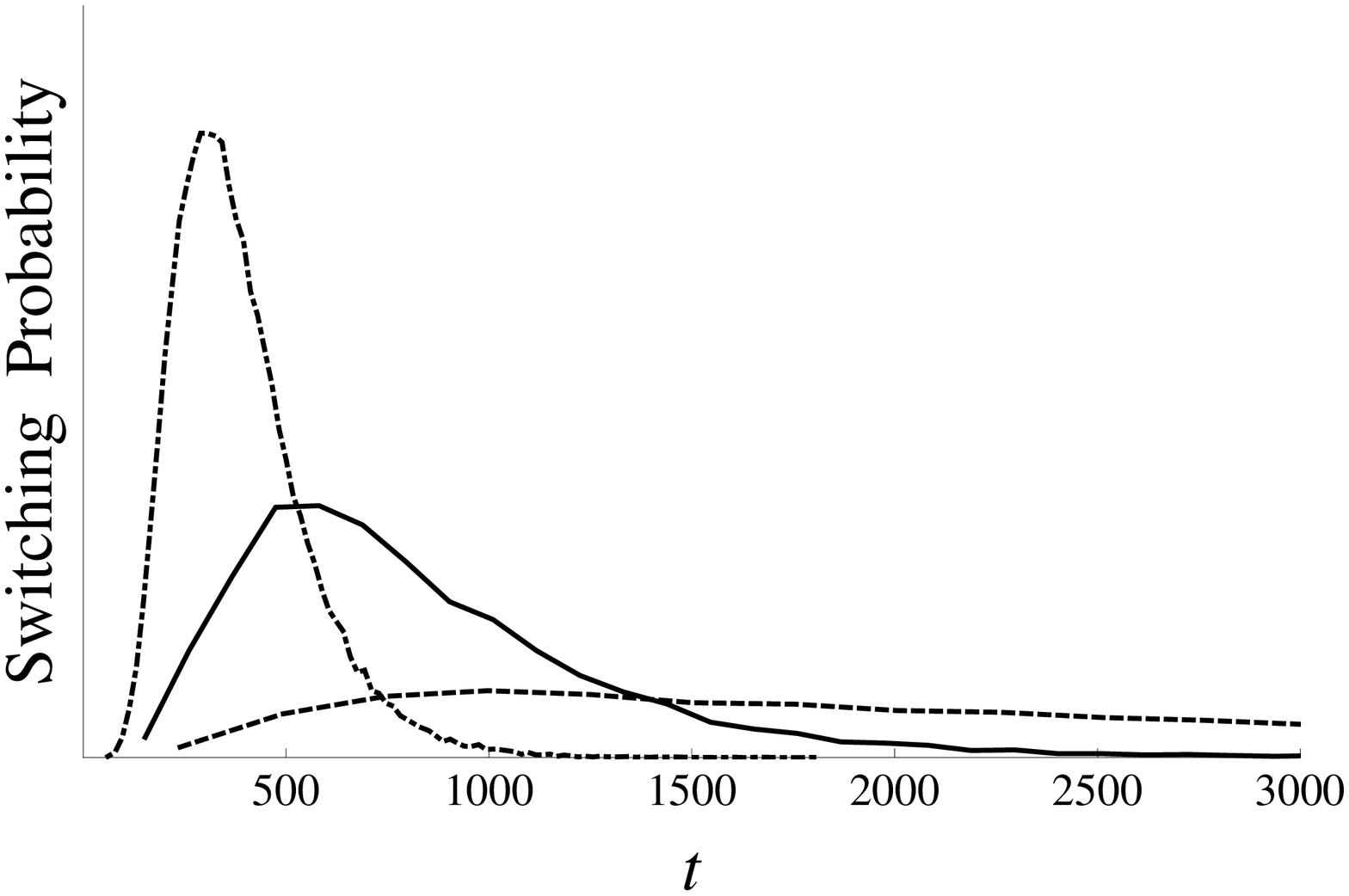}
\includegraphics[width=6cm]{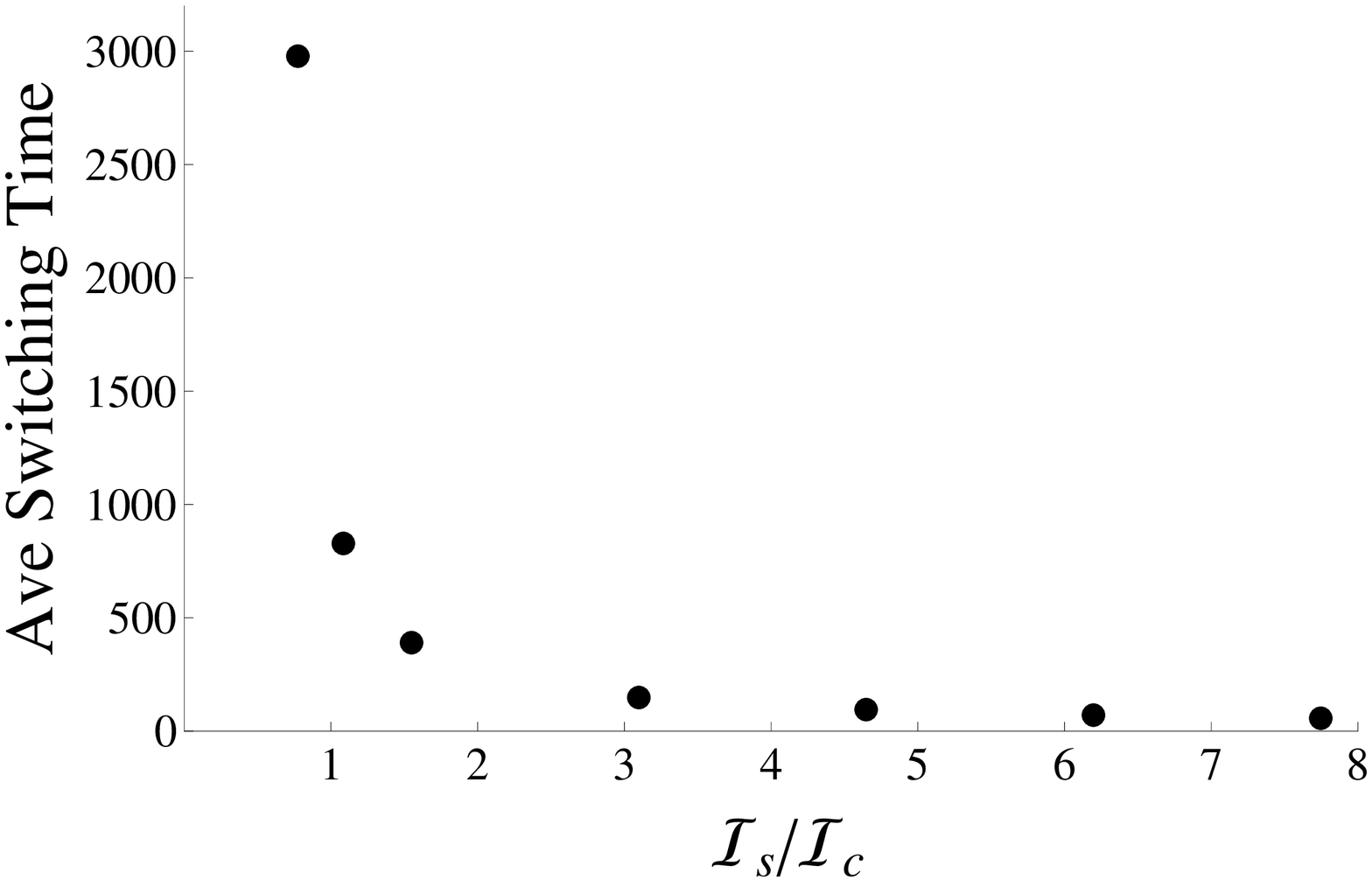}
\par\end{centering}
\caption{Numerical simulations of the switching process are shown for $H_k^z = 0.028 M_s$, $H_k^x = M_{\mathrm{s}}$, and $T=300 K$. (Left) Switching probability, as a function of time for spin-current values $1.6\,\mathcal{I}_{c}$ (dot-dashed), $1.1\,\mathcal{I}_{c}$ (solid), and $0.8\, \mathcal{I}_{c}$ (dashed). (Right) Average switching time as a function of spin-current relative to the the critical current, $\mathcal{I}_c$. Time is measured in units of $(\gamma M_\mathrm{s})^{-1}$}
\label{fig:switch}
\end{figure}

According to the deterministic equation of motion (\ref{eq:energy-rate}) one may introduce the {\em energy-dependent} critical spin-current
\begin{equation}
\mathcal{I}_c(E) =  \frac{ {F}_E}{ \mathbf{\hat z} \cdot {\mathbf{V}}_E}\,,
                                                                                 \label{eq:I-critical}
\end{equation}
which nullifies the right hand side of eqn~(\ref{eq:energy-rate}). The critical spin-current is then defined as
$  \mathcal{I}_c =\mbox{max}\{ \mathcal{I}_c(E)\,| \, 0<E<E_0\}$, so if the applied spin-current $\mathcal{I}_s>\mathcal{I}_c$ it forces
the magnetization direction to switch to a new stable energy minimum. The corresponding deterministic switching time is given by
\begin{eqnarray}
t_{\mathrm{sw}}(E_\mathrm{ini}) = \int^{E_0}_{E_\mathrm{ini}}  \frac{dE}{ \mathbf{\hat z} \cdot {\mathbf{V}}_E\left( \mathcal{I}_s - \mathcal{I}_c(E)\right) }\,.
\label{eq:Tsw}
\end{eqnarray}
where $E_\mathrm{ini}$ is an initial energy. This expression diverges as $ (\mathcal{I}_s - \mathcal{I}_c)^{-1}$ and such a tendency was indeed
observed in experiment \shortcite{Bedau10}. This divergence is augmented by the stochasticity, since even for the currents somewhat less than critical the switching  does occur, albeit taking exponentially long waiting time.   The proper description of the switching time must therefore
rely on the probability distribution to undergo the irreversible switch during time $t_\mathrm{sw}$. Such probability distributions calculated by Monte Carlo
simulations of eqn~(\ref{eq:LLG}) are presented in Fig.~\ref{fig:switch}a for several values of the spin-current $\mathcal{I}_s$. Having such
distributions, one may evaluate e.g. the average switching time,  plotted in Fig.~\ref{fig:switch}b as a function of the applied spin-current.
One observes that the switching time indeed grows exponentially at $ \mathcal{I}_s < \mathcal{I}_c$. Analytically this latter regime was analyzed using eqn~(\ref{eq:FP_general-current}) by Apalkov and Vissher \shortcite{ApalkovPRB}. Although Ref.~\shortcite{ApalkovPRB} used the
equilibrium noise, we found \shortcite{Dunn10} that inclusion of the shot-noise (\ref{eq:DI}) merely changes the effective temperature and therefore is hardly distinguishable from effects of  heating.

One may notice that eqn~(\ref{eq:Tsw}) for the deterministic switching time also tends to diverge if  $E_\mathrm{ini} \to 0$. This is due to the fact that if the initial direction of the magnetization is very close to the magnetic energy minimum, it formally takes a logarithmically long time to depart from the minimum. Clearly such a spurious divergency disappears once the distribution function of the initial (i.e. prior to the application of the spin-current) magnetic energy is taken into account. It is reasonable to assume that such a distribution is thermal and is given by
 eqn~(\ref{Boltsmann}) $\mathcal{P}_0(E_\mathrm{ini})\propto P_{E_\mathrm{ini}}  \exp\left[- E_\mathrm{ini}/k_\mathrm{B}T
\right]$. Although the (non-equilibrium) noise acts on the magnetization during the entire switching trajectory, we found that the simulation
data for $ \mathcal{I}_s > \mathcal{I}_c$, Fig.~\ref{fig:switch}, can be rather well fit by taking into account this initial distribution
only; i.e. assuming the deterministic motion with the random initial condition $\rho(t_{\mathrm{sw}}) = \int dE_\mathrm{ini}\,   \mathcal{P}_0(E_\mathrm{ini}) \delta(t_{\mathrm{sw}} - t_{\mathrm{sw}}(E_\mathrm{ini}))$. We therefore conclude that measuring the switching time distribution,  Fig.~\ref{fig:switch}, is not a sensitive way to resolve the effects of the non-equilibrium spin shot noise.

The treatment of the stochastic dynamics based on the time scales separation, presented in the previous section, is nevertheless very useful to
analyze an optimal protocol for the time-dependent spin-current pulse  $ \mathcal{I}_s (t)$.  The issue at hand is as follows: as shown above, to make the switching time shorter one needs a larger spin-current. This leads to an excessive Joule losses $\propto \int dt\, \mathcal{I}_s^{\,2}(t)$. On the other hand, decreasing the spin-current leads to a longer switching times, which in turn may again increase heating due to the long time needed to complete the switch. One may expect that there is an optimal strategy, which minimizes the heating losses. To find such a strategy  we treat the system as a heat engine and require that the ratio of the instantaneous energy gain $\dot E(t)$ to the energy loss $\propto \mathcal{I}_s^2(t) $ is maximized at each instance of time. Substituting the energy equation of motion (\ref{eq:energy-rate}) and taking variation with respect to the spin-current, we find the following condition
\begin{eqnarray}
0 = \frac{\delta}{\delta \mathcal{I}_s} \left(-\frac{{F}_E}{\mathcal{I}_s^2} + \frac{\mathbf{\hat z} \cdot {\mathbf{V}}_E}{\mathcal{I}_s} \right) = 2\frac{{F}_E}{\mathcal{I}_s^3} - \frac{\mathbf{\hat z} \cdot {\mathbf{V}}_E}{\mathcal{I}_s^2} \,.
\end{eqnarray}
Solving for the spin current $\mathcal{I}_s$, yields
\begin{eqnarray}
\mathcal{I}_s^{\mathrm{opt}} = 2\, \frac{{F}_E}{\mathbf{\hat z} \cdot {\mathbf{V}}_E} =2\, \mathcal{I}_c(E) \, ,
\label{eq:efficient-path}
\end{eqnarray}
cf. eqn~(\ref{eq:I-critical}).
Therefore the optimal current is exactly {\em twice} the one needed to nullify the energy flow in eqn (\ref{eq:energy-rate}). Substituting this back into eqn (\ref{eq:energy-rate}), one finds that upon the optimal current the latter takes the form
\begin{eqnarray}
\label{eq:time-reversed}
\dot{E} =   {F}_E\,.
\end{eqnarray}
This is exactly the same as in eqn~(\ref{eq:E_det0}) {\em without} the external spin current $\mathcal{I}_s$, but with the time being {\em reversed}. Therefore {\em the optimal current protocol is such that it exactly time-reverses the purely relaxational trajectory of an isolated system}. Such a relaxation trajectory must be thought of as starting at the energy maximum $E_0$ (i.e. the switching point) and winding down towards the energy minimum at $E=0$. This is a rather general statement which is based only on the assumption that the energy loss is quadratic in the spin-current (the latter is very nearly true practically in all systems investigated so far).

\begin{figure}[h]
  \begin{centering}
  \includegraphics[width=8cm]{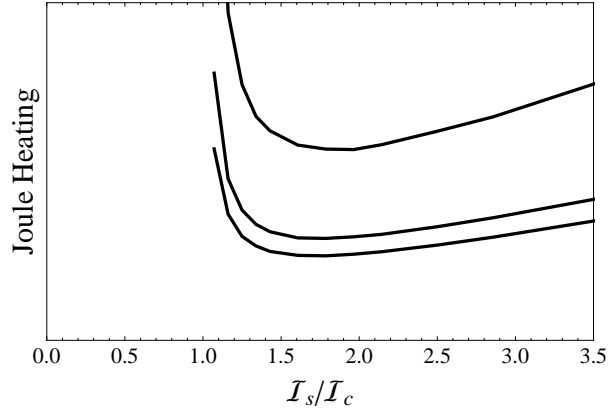}
  \par\end{centering}
  \caption{Simulated Joule heating during the switching process in arbitrary units vs. spin current for various confidence levels $p$. From top to bottom $p=0.99$, $p=0.5$, $p=0.2$. Here $\alpha = 0.01$, $H^z_k = 0.028 M_{\mathrm{s}}$, $H^x_k = M_{\mathrm{s}}$ and $E_{0}=140 k_B T$.}\label{fig:switching-energy}
\end{figure}

To put this observation on a more practical level we notice that the energy-dependent critical current (\ref{eq:I-critical}) appears to be  roughly energy independent.
Equation (\ref{eq:efficient-path}) implies then that the optimal current is very nearly a constant given by {\em twice} the critical switching current $\mathcal{I}_c$. This is shown in Fig.~\ref{fig:switching-energy} via Monte Carlo simulations of the Joule heating as a function of the spin current for various confidence levels.  At low confidence levels $\mathcal{I}_s^{\mathrm{opt}} \lesssim 2 \mathcal{I}_c$ due to large fluctuations, however at confidence levels nearing $100 \%$ we see the predicted $\mathcal{I}_s^{\mathrm{opt}} \approx 2 \mathcal{I}_c$.

\begin{figure}[h]
  \begin{centering}
  \includegraphics[width=8cm]{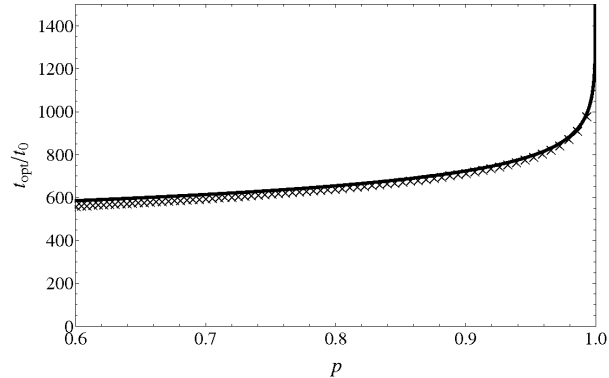}
  \par\end{centering}
  \caption{Simulated optimal switching time as a function of the confidence level (crosses), eqn~(\ref{eq:optimal-time}) (full line). Here $\alpha = 0.01$, $H^z_k = 0.028M_{\mathrm{s}}$, $H^x_k = M_{\mathrm{s}}$, $E_{0}=140 k_B T$ and $\mathcal{I}_s = \mathcal{I}_s^{\mathrm{opt}}$.}\label{fig:switching-time}
\end{figure}

While we have established that the optimal current pulse is approximately rectangular with the amplitude $\mathcal{I}_s^{\mathrm{opt}} \approx 2 \mathcal{I}_c$, we have not discussed yet its duration. As discussed above the distribution of the switching times is well described by the equilibrium thermal distribution of initial energy $\propto e^{-E_{\mathrm{ini}}/k_\mathrm{B}T}$ and the deterministic duration (\ref{eq:Tsw}).
Thus the times needed to reach the switching point, which are $\propto \alpha^{-1} \ln (E_0/E_{\mathrm{ini}})$, where $E_{0}=\mu_0 H_k^zM_\mathrm{s}/2$ is the hight of the energy barrier, are also scattered accordingly.    As a result, any pulse of a finite duration achieves switching only with a certain probability $p<1$. If this probability, i.e. error tolerance, is specified such that $1-p\ll 1$, all realizations with $E_{\mathrm{ini}}\geq k_\mathrm{B}T(1-p)$ should undergo the switch. This dictates that the duration of the optimal current pulse scales as
\begin{equation}\label{eq:optimal-time}
    t_{\mathrm{opt}} \approx \frac{t_0}{ \alpha}\,  \ln \left(\frac{E_{0}}{k_\mathrm{B} T(1-p)}\right)\,,
\end{equation}
where $t_0 = \left( \gamma M_\mathrm{s} \right)^{-1}$. This result is shown in Fig.~\ref{fig:switching-time} for higher switching probabilities and is in a good  agreement with the simulated data.

One can imagine sufficiently complex switching protocols, utilizing the precessional frequency to achieve  an optimal path. For example, Ref.~\shortcite{Cui2008} used current pulses resonant with the orbit period to excite the system. In the realm of relatively low frequency protocols, where the averaging over the fast angle is justified, twice the critical current pulse provides the best approximation to the optimal strategy.  The duration of this pulse weakly depends on the confidence level and can be estimated according to eqn~(\ref{eq:optimal-time}).

\section{Spectral width of the steady state  precession}
\label{sec:steady-state}

Another important application of spin-torque devices is provided by spin-torque oscillators. These devices are similar to spin valves, discussed above, but rather rely on the steady state  precession to generate microwave signal \shortcite{Deac08,Mizukami01,Krivorotov05,Nazarov06,Pribiag09,Georges09,Ruotolo09,Krivorotov2010,Krivorotov_cm1004,Rippard10,Krivorotov_cm1103}. Because of the high precision needed in  these applications it is important to understand how the noise can impact spectral line shape of the oscillator. There is a number of experimental
measurements of the irradiated frequency spectrum, which indicate a non-monotonous dependence of the spectral width on various parameters such as
the spin-current amplitude  \shortcite{Kiselev03,Mistral06,Slavin07,Georges09}.

The steady state magnetization precession (SSMP) occurs along SW orbit when the energy pumped by  the spin-torque is compensated by the dissipation. If this is the case, the time derivative of the total energy along a certain orbit, eqn~(\ref{eq:energy-rate}), vanishes, $\dot{E}=0$ at $E=\bar E$, full line in Fig.~\ref{fig:energycurrent}.
Generically, SSMP is realized within a certain range of the spin-currents  below the critical one   $\mathcal{I}_s \lesssim \mathcal{I}_c$.
The corresponding energy $\bar E=\bar E(\mathcal{I}_s)$ of the stable SW orbit typically runs from zero to the switching energy $E_0$, with the
increase of the spin-current.
The energy distribution corresponding to the steady state magnetization precession can be obtained from  eqn~(\ref{eq:stat}). As follows from eqn (\ref{eq:energy-rate}) the generalized force
$\dot{E}=-  F_{\bar{E}} + {\bf \hat{\mathcal{I}}_s}\cdot \mathbf{V}_{\bar{E}}=0$, thus the energy distribution can be obtained by expanding the  force as $\dot{E}\approx \frac{d\dot{E}(\bar{E})}{dE}(E-\bar{E})$. Then, employing eqn~(\ref{eq:stat}) and neglecting the energy dependence of the diffusion coefficient, one obtains
\begin{equation}
\mathcal{P}^{\mathrm{SSMP}}_0(E)\propto \exp\left[\frac{d\dot{E}(\bar{E})/dE}{2\mathcal{D}_{\bar{E}}} \left(E-\bar{E}\right)^2\right].
\label{rho_SSMP}
\end{equation}
Therefore, the  SSMP energy distribution is approximately Gaussian, centered at the energy $\bar{E}$ with the width determined by the diffusion coefficient and the derivative of the generalized force $\dot{E}$ over the energy.

To investigate the spectral width of the precession spectrum, one needs to restore the stochastic equation (\ref{phi_stoch}) for the angular variable $\varphi(t)$ running along SW orbits. Employing the fact that the correlator (\ref{eq:DI}) of the stochastic magnetic fields is isotropic, it may be written as
\begin{equation}
\dot{\varphi}= \Omega_{E}  + |{\bf g}_{\varphi}(E,\varphi)| \xi_{\varphi}(t) \,; \quad\quad\quad
|{\bf g}_{\varphi}(E,\varphi)|=\frac{2\pi}{P_{{E}}}\, \frac{M_\mathrm{s}}{|[{\bf M}\times{\bf H}_\mathrm{eff}]|}\, ,
\label{eq:varphi}
\end{equation}
where $\xi_\varphi(t)$ is a {\em scalar} Langevin force with the correlator
$\langle \xi_{\varphi}(t)\xi_{\varphi}(t')\rangle= 2D({\bf M})\delta(t-t') $.
It is thus the function $|{\bf g}_{\varphi}(E,\varphi)|$, which determines the  frequency spectrum.  Some of it universal features can be inferred without relying on a specific model. In particular, it can be concluded that the function $|{\bf g}_{\varphi}(E,\varphi)|$ diverges as $E$ tends to zero. Indeed, at small energies $E$, the characteristic size of  SW orbit goes to zero (at zero energy the orbit shrinks to a point), while the influence of noise on the change of angle $\varphi$ grows.  The character of the singularity  can be obtained by considering the simplest model of steady state precession in a system with an easy axis anisotropy and both external magnetic field and spin current being parallel to the easy axis. In this case the energy is a function of $\cos\theta$ and SW orbits  are circles, while $\varphi$ is the azimuthal angle and $|{\bf g}_{\varphi}(E,\varphi)|\propto 1/\sin\theta$ \shortcite{Chudnovskiy08}. Taking into account that the energy minimum corresponds to $\theta=0$, and expanding at small angles  $E(\cos\theta)\propto \theta^2$, one concludes that $|{\bf g}_{\varphi}(E)|\propto 1/\sqrt{E}$.
Therefore, at small energies of SW orbits, the noise strength decreases with the  growth of the energy. This feature is responsible for the decreasing linewidth of the microwave spectral power with the increase of the spin-current (and thus increase of $ E$), as discussed in Refs.~\shortcite{Slavin07,Kim07} assuming  the equilibrium thermal noise \shortcite{Brown63}.

To evaluate the spectral power, one notices that the precessing magnetization induces the oscillating magnetic field ${\mathcal H}$ in a waveguide, which is a periodic function of the angle $\varphi$, i.e ${\mathcal H}(\varphi) ={\mathcal H}(\varphi+2\pi)$. The spectral power   is given by
$S(\omega)= \int dt  \left\langle {\mathcal H}(\varphi(t)) {\mathcal H}(\varphi(0))\right\rangle_{\xi_\phi} e^{-i\omega t}$ and according
to eqn~(\ref{eq:varphi}) it consists of series of peaks at multiple integers of $\Omega_{\bar E}$. Focusing on the linewidth of the $n$-th harmonics $\omega= n\Omega_{\bar E}$, one needs to calculate
\begin{equation}
S(\omega)\propto  \int dt  \left\langle e^{in\left[\varphi(t) - \varphi(0)\right]}\right\rangle_{\xi_\phi} e^{-i\omega t}\,; \quad\quad \quad
\varphi(t)=\Omega_E t +\int^t |{\bf g}_{\varphi}(E,\varphi)|\,\xi_{\varphi}(t')dt' \, ,
\label{S_as_phiphi}
\end{equation}
where we have employed the formal solution of eqn~(\ref{eq:varphi}).
The averaging over the noise $\xi_\phi$ is thus reduced to the Gaussian
integral, which gives
\begin{equation}
S(\omega) \propto \int dt  \,\exp\left\{i(n\Omega_E-\omega)t - \frac{n^2}{2} \int_{0}^{t}D({\bf M}(t'))\,  |{\bf g}_{\varphi}(E,\varphi(t'))|^2 dt' \right\}.
\label{S_as_exp}
\end{equation}
This expression should be now averaged over the stationary energy
distribution (\ref{rho_SSMP}). Below we shall assume that the
latter is rather narrow and forces the energy to be close to the
stationary value $\bar E$. In this case the spectrum is Lorentzian
centered around $\omega=n\Omega_{\bar E}$ with the width given by
$n^2 \Delta_{\bar E}$, where $\Delta_{\bar E}$ is the zeroth
Fourier harmonics of the last term in the exponent in
eqn~(\ref{S_as_exp})
\begin{equation}
\Delta_{\bar E}=\frac{1}{2\pi }\int_0^{2\pi}\! D({\bf M})\, |{\bf
g}_{\varphi}(\bar{E}, \varphi)|^2\, d\varphi = \frac{2\pi
}{P_{\bar{E}}^2} \int_0^{2\pi}    \frac{M^2_\mathrm{s} D({\bf M})
}{\left|[{\bf M}\times{\bf H}_{\mathrm{eff}}]\right|^2}\, d\varphi
. \label{Spectral_Width}
\end{equation}

\begin{figure}
\begin{centering}
\includegraphics[width=8cm,height=7cm,angle=0]{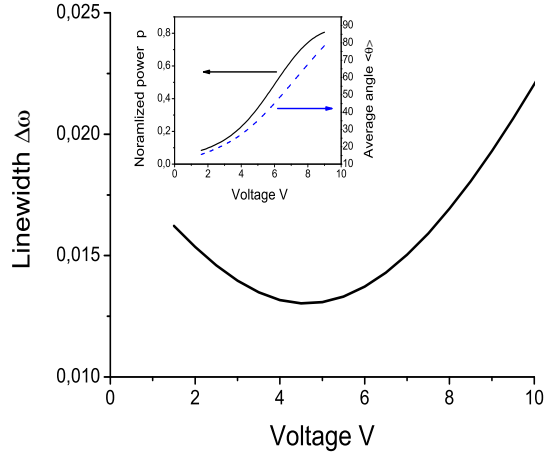}%
\par\end{centering}
\caption{Calculated linewidth of the microwave spectral
power vs. applied voltage in units of temperature.
Inset: calculated microwave power (solid line) and the average
precession angle  (dashed line) vs. voltage.
Parameters: $M_\mathrm{s}/(\hbar \gamma)=10$, $\hbar\gamma H^z_k/k_B=3$K, $T=1$K,
$\alpha_0=0.01$.  Conductances in units $e^2/h$: $G_P=0.181$, $G_{AP}=0.019$.
$d{\bf I}_\mathrm{sf}/dV=0.01 e$.
\label{fig1}}
\end{figure}

Finally, using the equation of motion (\ref{eq:precession}), one may rewrite the spectral width as an integral along the static SW orbit with the energy $\bar E$
\begin{equation}
\Delta_E =\frac{4\pi^2 M^2_\mathrm{s}}{\gamma P_{\bar E}^3} \oint
D({\bf M})\, \frac{\left[d\mathbf{M}\times
\mathbf{H}_{\mathrm{eff}}\right] \cdot \mathbf{M}} {\left|[{\bf
M}\times{\bf H}_{\mathrm{eff}}]\right|^4} , \label{eq:Delta}
\end{equation}
compare this result with that for the energy diffusion coefficient (\ref{D_E}).
Proportionality of the spectral width to the noise correllator $D({\bf M})$, eqn~(\ref{eq:DI}), implies that it is proportional to
temperature if $T>eV$. On the other hand, at small temperature and/or larger voltage $V$ the correlator is dominated by the non-equilibrium noise and one expects $\Delta\propto V\propto \mathcal{I}_s $. Therefore one expects decrease of the spectral width with decreasing temperature and saturation on  a voltage-dependent level \shortcite{Ralph05}.

Figure \ref{fig1} shows the calculated spectral linewidth as a function of the applied voltage at $T=1K$. Notice the non-monotonous behavior, similar to the one observed in a number of experiments  \shortcite{Rippard04,Ralph05,Rippard06,Mistral06,Slavin07}.
The initial decrease of the spectral width is associated with the angular factor $\left|{\bf M}\times{\bf H}_{\mathrm{eff}}\right|^2\propto \bar E$   in the denominator of eqn~(\ref{Spectral_Width}) and the fact that the stable energy $\bar E$ grows with the applied spin-current. On the other hand,
the subsequent reversal of the trend and growth of the spectral width is due to the fact that $D({\bf M})$ grows with the voltage, because of the   spin-torque shot noise, eqn~(\ref{eq:DI}).

\section{Conclusion and acknowledgments}

Miniaturization of spintronics devices puts the stochastic effects in the magnetization dynamics on the cutting edge of investigations. While the fundamentals of stochastic description of magnetization in terms of the FP equation have been developed decades ago, the highly nonequilibrium regimes used in modern devices require a specific description of non-equilibrium noise. In this work, a step towards such a description has been reviewed. As we showed, despite a large variety of dynamical regimes, they all have in common the time scale  separation between the fast and almost energy conserving magnetization precession and the slow energy change. This point allows to develop a theoretical description of magnetization dynamics in terms of slowly varying energy of the precessional orbit. The resulting FP equation is one dimensional, which greatly simplifies its treatment in comparison to the initial FP equation for the magnetization vector. In this work we applied such an approach to the analysis of current induced magnetization switching and steady state magnetization precession, obtaining results on the optimization of the current switching protocol and on the linewidth of radiation emitted by steady state magnetization precession.
Further development of the energetic description is required by the dynamical regimes, where resonances between the precessional orbits with different energies are induced by applying an ac-modulated spin current as reported recently in e.g. \shortcite{Krivorotov2010}, as well as for the description of spin switching by spin current pulses of complex form \shortcite{Cui2008}.


We are grateful to  Ilya Krivorotov for fruitful discussions.
ALC acknowledges support from DFG through the Collaborative Research Center 668.
AK and TD were supported by NSF
Grant DMR-0804266 and U.S.-Israel Binational Science
Foundation Grant 2008075.

\bibliographystyle{OUPnamed_notitle}
\bibliography{bibliography}

\end{document}